\documentclass[aip,reprint]{aip/revtex4-1}

\usepackage[utf8]{inputenc} 
\usepackage[T1]{fontenc}    
\usepackage{url}            
\usepackage{booktabs}       
\usepackage{amsfonts}       
\usepackage{nicefrac}       
\usepackage{microtype}      
\usepackage{graphicx}
\usepackage{lipsum}
\usepackage{comment}
\usepackage[caption=false]{subfig}
\usepackage{textcomp}
\usepackage{soul}           
\usepackage{amsmath}
\usepackage{listings}
\usepackage{chngcntr}
\usepackage{enumitem}
\usepackage[table,xcdraw]{xcolor}
\usepackage{booktabs}
\usepackage{multirow}
\usepackage{hhline}
\usepackage[english]{babel}
\usepackage{siunitx}
\usepackage{setspace}

\usepackage{amsmath,amsfonts,amssymb}

\usepackage{url}            
\usepackage{booktabs}       
\usepackage{amsfonts}       
\usepackage{nicefrac}       
\usepackage{microtype}      
\usepackage{xcolor}   
\usepackage{graphicx}
\usepackage{soul}
\usepackage{bm}
\usepackage{placeins}
\usepackage{booktabs}
\usepackage{multirow}
\usepackage{hyperref}       
\usepackage{cleveref}

\newcommand{\msf}[1]{\mathsf{#1}}

\newcommand{\R}{{\mathbb{R}}}

\crefname{assumption}{Asm.}{Asm.}
\crefname{equation}{}{}
\Crefname{equation}{Eq.}{Equations}
\crefname{figure}{Fig.}{Figs.}
\crefname{table}{Tab.}{Tabs.}
\crefname{section}{Sec.}{Sec.}
\crefname{theorem}{Thm.}{Thm.}
\crefname{lemma}{Lemma}{Lemmas}
\crefname{corollary}{Cor.}{Cor.}
\crefname{example}{Example}{Examples}
\crefname{remark}{Remark}{Remarks}
\crefname{algorithm}{Alg.}{Algorithms}
\crefname{appendix}{Appendix}{Appendices}
\crefname{subappendix}{Appendix}{Appendices}
\crefname{subsubappendix}{Appendix}{Appendices}

\newcommand{\algo}{\texttt{franken}}

\newcommand{\paragraphtitle}[1]{\textsf{\textbf{\small{#1}}}}
\draft 

\begin{document}

\title{Fast and Fourier Features for Transfer Learning of Interatomic Potentials}

\author{Pietro Novelli}
\affiliation{Computational Statistics and Machine Learning, Italian Institute of Technology, Genova, Italy}

\author{Giacomo Meanti}
\affiliation{Centre Inria de l'Université Grenoble Alpes, Montbonnot, France}

\author{Pedro J. Buigues}
\affiliation{Computational Statistics and Machine Learning, Italian Institute of Technology, Genova, Italy}
\affiliation{Atomistic Simulations, Italian Institute of Technology, Genova, Italy}

\author{Lorenzo Rosasco}
\affiliation{MaLGa Center - DIBRIS, Università di Genova, Genoa, Italy}
\affiliation{Center for Brains, Minds and Machines, MIT, Cambridge, MA, USA}
\affiliation{Italian Institute of Technology, Genova, Italy}

\author{Michele Parrinello}
\affiliation{Atomistic Simulations, Italian Institute of Technology, Genova, Italy}

\author{Massimiliano Pontil}
\affiliation{Computational Statistics and Machine Learning, Italian Institute of Technology, Genova, Italy}
\affiliation{AI Centre, University College London, London, United Kingdom}

\author{Luigi Bonati}
\email[]{luigi.bonati@iit.it}
\affiliation{Atomistic Simulations, Italian Institute of Technology, Genova, Italy}

\begin{abstract}
Training machine learning interatomic potentials that are both computationally and data-efficient is a key challenge for enabling their routine use in atomistic simulations. To this effect, we introduce \algo, a scalable and lightweight transfer learning framework that extracts atomic descriptors from pre-trained graph neural networks and transfer them to new systems using random Fourier features—an efficient and scalable approximation of kernel methods. \algo~enables fast and accurate adaptation of general-purpose potentials to new systems or levels of quantum mechanical theory—without requiring hyperparameter tuning or architectural modifications. On a benchmark dataset of 27 transition metals, \algo~outperforms optimized kernel-based methods in both training time and accuracy, reducing model training from tens of hours to minutes on a single GPU. We further demonstrate the framework’s strong data-efficiency by training stable and accurate potentials for bulk water and the Pt(111)/water interface using just tens of training structures. Our open-source implementation (\url{https://franken.readthedocs.io}) offers a fast and practical solution for training potentials and deploying them for molecular dynamics simulations across diverse systems.
\end{abstract}

\maketitle

Molecular dynamics (MD) simulations are a powerful tool for obtaining atomistic insights into a wide range of processes, from material properties to chemical and catalytic reactions~\cite{frenkel2023understanding}. These simulations integrate the equations of motion using a model for atomic interactions --- the potential energy surface (PES). Traditionally, the PES is derived from quantum mechanical calculations such as density functional theory (DFT)~or from simplified empirical models fitted to experimental data or first-principles calculations. However, quantum mechanical methods, while accurate, are computationally expensive, and empirical force fields often lack transferability and accuracy. In recent years, machine learning interatomic potentials (MLIPs) have emerged as a compelling solution, striking a balance between the accuracy of DFT and the efficiency of empirical force fields. These models learn the PES via regression of large datasets of quantum mechanical calculations, thereby achieving quantum-level accuracy while enhancing the inference speed by orders of magnitude.

For extended systems, MLIPs are typically constructed expressing the total energy as a sum of local atomic contributions that depend on the surrounding environment. To ensure physical consistency, the PES must be invariant to relevant symmetries, such as roto-translation and permutations of atoms of the same chemical species.
Early MLIPs relied on handcrafted descriptors to encode the local atomic environments in a way that preserves physical symmetries. Two widely used approaches are Behler–Parrinello symmetry functions~\cite{behler2007generalized,behler2011atom}, and the Smooth Overlap of Atomic Positions (SOAP)~\cite{bartok2013representing}. Symmetry functions capture radial and angular correlations using predefined functional forms designed to respect rotational and translational symmetries. Although protocols to automatically select informative symmetry functions from large candidate pools exist~\cite{imbalzano2018automatic}, this procedure becomes increasingly difficult as the number of chemical species grows, since the distinct atomic interactions that must be described increase combinatorially with the number of species. On top of this, the model architecture should also be carefully optimized in order to achieve optimal results~\cite{kyvala2023optimizing}. SOAP represents the local atomic environment around a central atom using a smoothed atomic density projected onto a basis of radial and spherical harmonic functions. The overlap of these densities is then used to compare the environments, typically in a kernel-based~\cite{scholkopf2002learning, shawe2004kernel} framework such as Gaussian Approximation Potentials ~\cite{bartok2010gaussian}. This approach, however, suffers from two key limitations: its dimensionality grows quadratically with the number of chemical species, as separate channels are typically used for each pairwise element combination, while its three-body nature limits the accuracy in representing highly complex interactions.
More recently, the Atomic Cluster Expansion (ACE)~\cite{drautz2019ace} has been proposed as a systematic and general framework for constructing many-body descriptors. Compared to earlier approaches, ACE offers improved scalability and flexibility by organizing atomic interactions into a hierarchy of body-order terms and employing a shared basis across chemical species. This leads to a more favorable, typically linear, scaling with the number of species. However, the number of basis functions can still become large for high body orders or complex systems, and the method relies on careful hyperparameter tuning and sparsification to remain computationally tractable.

For the class of methods outlined above, training of MLIPs can be decomposed into two main tasks: selecting a suitable representation and regressing energy and forces as functions of that representation. Following the deep learning approach, methods such as DeepMD~\cite{zhang2018deep} are fully data-driven, using one neural network to learn the representation and another to predict target properties. Geometric graph neural networks (GNNs)~\cite{reiser2022graph}, based on message-passing schemes, go a step further by integrating representation learning directly into the model architecture. Atoms are treated as nodes in a graph, with edges connecting atoms whose pairwise distances fall within a specified cutoff. Node features are then updated iteratively by aggregating information from neighboring atoms. Repeating this process over multiple message-passing layers allows the model to construct expressive representations that capture complex many-body interactions and chemical environments, thereby tightly coupling representation learning and regression.
In terms of the scaling with the number of chemical species, GNNs typically encode chemical identity via a fixed-dimensional embedding at the node level. Hence, the representation size remains constant with respect to the number of species, making GNNs particularly appealing for modeling chemically diverse systems. Moreover, GNNs offer a natural framework for encoding physical symmetries, as permutation and translational invariance are built into the graph structure and aggregation operations. 
Furthermore, recent advances have enabled explicit rotational equivariance through specialized architectures~\cite{batatia2025design}. These allow modeling of vector and tensor-valued properties, and even for invariant quantities such as the potential energy, equivariant models have demonstrated improved data-efficiency and generalization~\cite{fu2023forces}. Altogether, these features make GNN-based MLIPs a powerful and flexible class of models, unifying representation learning and regression while incorporating the symmetries of atomic systems.

Despite these advances, constructing an MLIP for a specific system remains a non-trivial task. A critical challenge is assembling a training dataset that spans all the relevant phase space~\cite{kulichenko2024data}—an unattainable requirement through ab initio molecular dynamics alone. This challenge often requires the use of active learning strategies~\cite{jinnouchi2020fly,schran2020committee} to iteratively refine training datasets and enhanced sampling techniques to capture rare events. Such events include phase transitions~\cite{bonati2018silicon,abou2024unraveling} and chemical reactions~\cite{yang2022using,guan2023using,bonati2023role,david2025arcann} that occur on time scales inaccessible to unbiased simulations. For MLIPs to become routinely applicable, we also need models that ensure stable and physically accurate molecular dynamic simulations. Indeed, while the goodness of MLIPs is primarily evaluated according to the prediction error it attains on a held-out dataset, the ultimate goal is to produce reliable MD simulations~\cite{stocker2022robust,fu2023forces,guan2023using}. However, models with comparable force predictions on a test dataset can exhibit vastly different behaviors when deployed "in the wild". Recently, equivariant neural networks such as Nequip~\cite{batzner2023nequip}, Allegro~\cite{musaelian2023learning} or MACE~\cite{batatia2022mace} have shown to improve the data-efficiency for molecular dynamics, allowing stable simulations already when trained with a few hundreds of samples~\cite{fu2023forces}. 

A recent paradigm shift has occurred with the emergence of large datasets of DFT calculations for specific domains, such as OpenCatalyst~\cite{chanussot2021open,tran2023open,barroso_omat24} for catalysis and Materials Project~\cite{jain2013commentary,deng2023chgnet} or Alexandria for materials~\cite{Ghahremanpour_2018}. GNNs trained on these datasets such as MACE-MP0~\cite{batatia2023foundation}, SevenNet0~\cite{park2024scalable}, CHGNet~\cite{deng2023chgnet}, M3GNet~\cite{Chen_2022} and MatterSim~\cite{yang2024mattersim},introduced the paradigm of general-purpose or "universal" potentials. Unlike previous approaches, where each potential was typically trained and used only for a specific system, these large models - trained over millions of configurations - have proven adaptable, offering stable simulations even for systems not explicitly included in the training set. This has prompted many efforts in assessing their reliability in a range of zero-shot applications.\cite{P'ota_2024,Peng_2024,shiota2024tamingmultidomainfidelitydata,Wang_2025}.
Even if these large MLIPs provide stable and qualitatively correct molecular dynamics, they often lack quantitative reliability for predictive applications. This can occur for different reasons. For instance, the target system may lie outside the training distribution, leading to poor generalization. Additionally, the level of DFT theory used to generate the reference data may not be sufficiently accurate for a specific system, requiring additional fine-tuning to provide a physically meaningful model. These factors highlight the fundamental need for strategies to adapt large-scale MLIPs to specific systems. Fine-tuning the weights of a universal model is a natural approach, although the optimal strategy remains unclear. Options range from using pre-trained weights and continuing training on a new dataset,~\cite{P'ota_2024,kaur2024dataefficientfinetuningfoundationalmodels} selectively freezing model layers and optimizing the rest,~\cite{radova2025fine, deng2024overcomingsystematicsofteninguniversal} or employing multi-head architectures with replay training~\cite{radova2025fine}. The latter strategy requires training with (part of) the original dataset, resulting in a significant computational cost. 

\begin{figure*}[t!]  
    \centering
    \includegraphics[width=\linewidth]{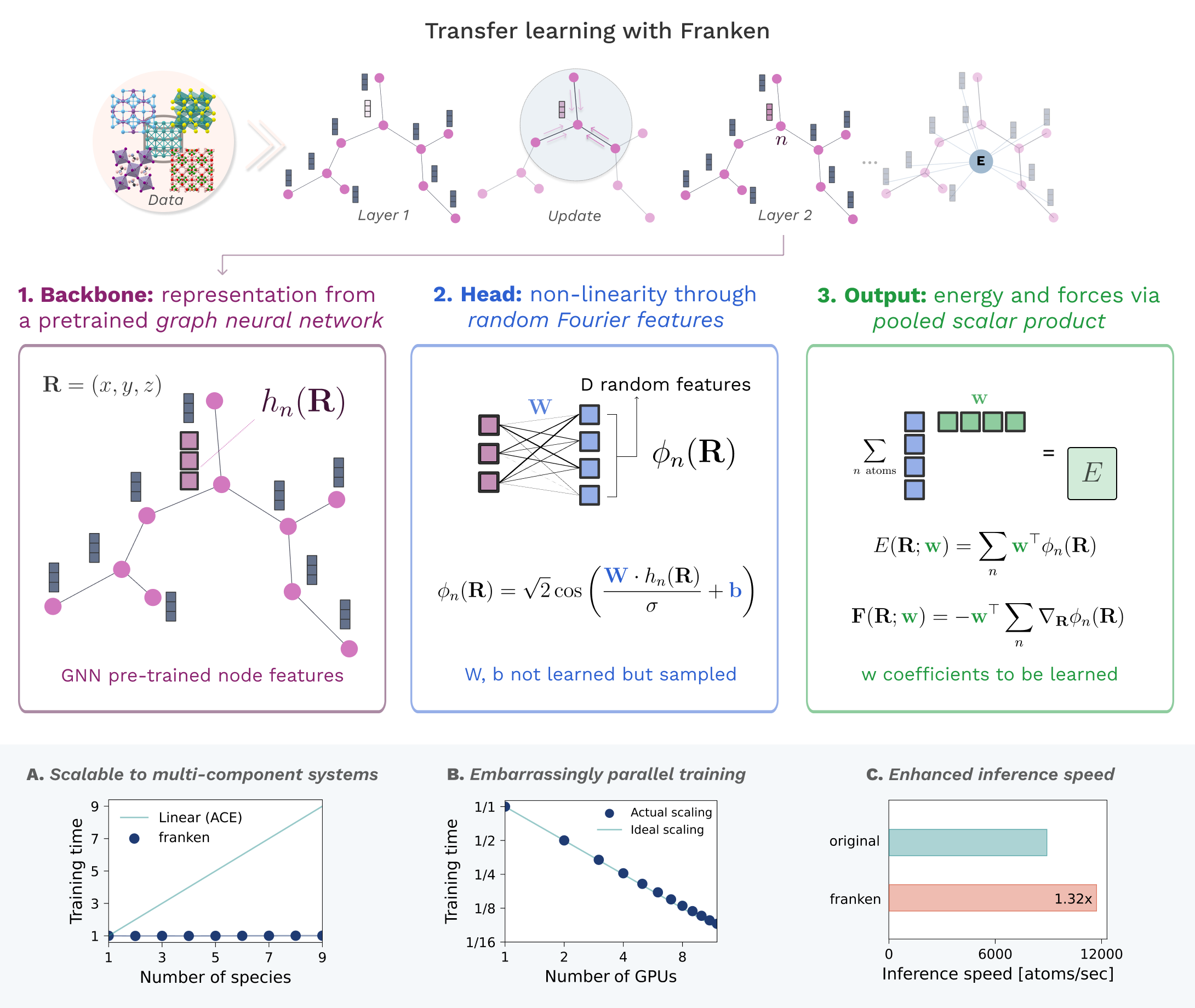}
    \caption{\paragraphtitle{The components of \algo}. 1) The backbone (left, red box) utilizes a pre-trained message-passing neural network to extract invariant descriptors $h_n(\mathbf{R})$ from atomic environments, where $\bm{R}$ represents atomic coordinates. These representations capture local chemical environments through pre-trained node features. 2) The head component (middle, blue box) introduces non-linearity through random Fourier features, transforming the GNN descriptors using randomly sampled weights $\bm{W}$ and bias $\bm{b}$ to generate features $\phi_n(\mathbf{R})$ via a cosine transformation. This random feature mapping approximates a Gaussian kernel. 3) The output module (right, green box) computes atomic energies through a learnable weighted sum of these features, with the total energy $E$ being the sum across all atoms. Forces $\bm{F}$ are calculated as the negative gradient of energy with respect to atomic positions. 
    In the bottom row, we highlight three properties enabled by~\algo. (A) Constant training times irrespective of the number of chemical species in the system (unlike methods based on physical descriptors, which scale at least linearly). (B) Near-ideal scalability of \algo~across multiple GPUs, with training time decreasing linearly as GPU count increases.  (C) Significantly faster inference speeds when compared to the original backbone GNN model, with up to 1.32$\times$ speedup (in the case of MACE-MP0-L0, see detailed timings in \cref{table:SI-timings}).}
    \label{fig:franken-diagram}
    
\end{figure*}

The question of how large MLIP models can be adapted to specific systems can be framed in the broader context of transfer learning~\cite{zhuang2020comprehensive}, in which information from one pre-trained model is exploited to accelerate the learning process of another one, such as in Refs. \citenum{chen2023data,zaverkin2023transfer,falk2024transfer,bocus2025operando,rocken2025enhancing}.
In our previous work~\cite{falk2024transfer}, we explored a transfer learning approach by first extracting intermediate representations from a pre-trained GNN on the OC20 dataset~\cite{chanussot2021open} and then using kernel mean embeddings~\cite{muandet2017kernel} to transfer this information to new systems. 
This approach provided a simple and efficient solution for learning the potential energy surface in a data-limited regime and also demonstrated good transferability. 
However, kernel methods suffer from poor scalability in both training and inference, making this method impractical for MD simulations.
To overcome these limitations, one can resort to large-scale kernel techniques. For instance, the Nystr{\"o}m approximation~\cite{williams2000nystrom,rudi2015nystrom} allows one to scale to very large datasets~\cite{meanti2020falkon} by approximating the kernel matrix through low-rank decompositions. Alternatively, random features~\cite{rahimi2007random} (RFs) efficiently approximate the kernel by projecting the infinite-dimensional kernel map into a randomized finite-dimensional feature space, effectively turning kernel-based learning into a linear regression on the random features. In the specific case of MLIPs, the scalar potential can be learned alongside the forces~\cite{shi_2010}, exploiting automatic differentiation.
Both RFs and Nystr{\"o}m have been leveraged to accelerate the training and inference time of kernel-based interatomic potentials~\cite{bartok2010gaussian,dhaliwal2022machine}. However, the scalability of such approaches remains constrained by the reliance on explicit physical descriptors, which can become computationally demanding as the complexity of the system increases.

In this manuscript, we build on these ideas and propose \algo, an extremely efficient and scalable transfer learning framework to build MD-capable machine learning potentials. \algo~efficiently adapts the learned representations to new systems by combining pre-trained GNN embeddings with random features. Through extensive experiments on different systems, we demonstrate that \algo~achieves accuracy comparable to MLIPs based on physical descriptors, eliminates the need for extensive hyperparameter tuning, and enables stable and data-efficient MD simulations with as little as tens of training samples.

\section*{Results}

\subsection*{The transfer learning algorithm.}
\vspace{-1em}
\begin{figure*}[t!]  
    \centering
    \includegraphics[width=0.85\linewidth]{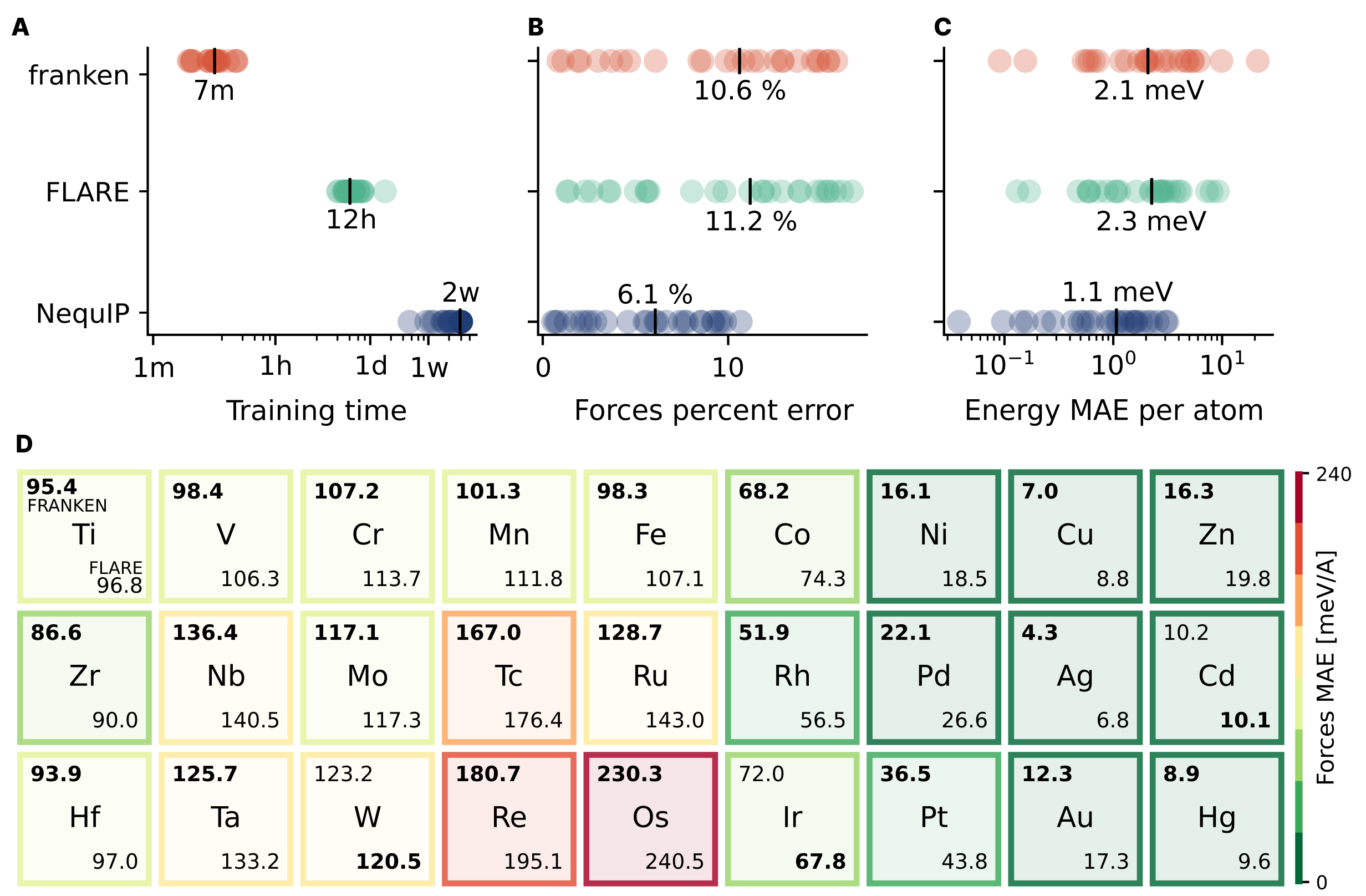}
    \caption{ \paragraphtitle{Accuracy vs efficiency on the TM23 dataset}. (A) Train time across the TM23 dataset;  \algo~MLIPs require only minutes for training compared to hours (FLARE) or weeks (NequIP). (B) Force percent error, as defined in~\citenum{owen2024complexity}, for each element. (C) Energy errors across the dataset.
    (D) Accuracy across the periodic table. Mean absolute force error across the 27 transition metals of TM23~\cite{owen2024complexity} of \algo~with MACE-MP0 backbone and FLARE, a kernel method using physical descriptors. For each element, the \algo's MAE is reported on the top left, while  FLARE's one on the bottom right; the minimum one is indicated in bold with a darker background. FLARE's results are taken from Ref.~\citenum{owen2024complexity} and refer to the ACE B2 descriptors (three-body representation), whose parameters (angular and radial fidelity, cutoff) have been optimized for each different elemental system. \algo~generally attains more accurate forces compared to FLARE B2, with an average $8.9\%$ lower MAE across chemical species, while requiring no selection of physical parameters, providing an accurate representation across the periodic table.
    }
    \label{fig:TM23-efficient}
\end{figure*}

\algo's MLIPs are created via "model surgery" between the inner layers of a pre-trained GNN and additional yet-to-train neural network blocks. Our open-source code executes this procedure seamlessly, and the end-user will work with a single final MLIP, which can be easily deployed to perform molecular dynamics simulations. In particular, the architecture of \algo~can be decomposed into three functionally different blocks (\cref{fig:franken-diagram}). In the following, we highlight their main features, referring to the methods for a more detailed description.

The first block is responsible for extracting the representation of the chemical environment for each atom in a given input configuration $\bm{R}$. Specifically, the $n$-th atom is associated with a vector $h_{n}(\bm{R})$ of $\text{SO}(3)$-invariant node-level descriptors from the inner layers of a pre-trained GNN (see Methods). This vector provides an effective representation of the local environment that respects relevant symmetries, with a computational and memory footprint not depending on the number of chemical species, yielding an efficient strategy even for multicomponent systems (\cref{fig:franken-diagram}A).

The second block of \algo~models is the core of our transfer learning approach, which transforms the pre-trained GNN descriptors $h_{n}$ into the final features used to predict energy and forces. To this end, we employ random Fourier features\cite{rahimi2007random, pham2013fast, yu2016orthogonal} (RF) maps $\phi_{n}(\bm{R}) = \phi(h_{n}(\bm{R}))$, a class of transformations introduced to scale kernel methods to large datasets while still preserving their key learning guarantees~\cite{rudi2017generalization, mei2022generalization}. Specifically, random features approximate a kernel function $k$ as a dot product in a suitable feature space: $$k(h, h') \approx \phi(h)^{\top} \phi(h').$$
In \algo~we use the trigonometric RF map (Eq.~\ref{eq:rfs} and ~\cref{fig:franken-diagram}) that approximates the Gaussian kernel~\cite{rahimi2007random,yu2016orthogonal} with a given length-scale (see Methods). Notably, at variance with other kernel approximation techniques\cite{rudi2015nystrom}, the RF maps can be implemented as a simple neural network layer with cosine nonlinearity, making its integration into existing deep-learning pipelines extremely simple. 

Thanks to this approach, the atomic energy for the $n$-th atom can be simply modeled as the scalar product between the RFs $\phi_{n}(\bm{R})$, and a {\em learnable} vector of coefficients
$E_{n}(\bm{R}; \bm{w}) =  \bm{w}^{\top} \phi_{n}(\bm{R})$. The total energy is obtained by summing together the atomic contributions, while the forces are calculated as the gradient of the total energy, making it a conservative force field.
As detailed in the Methods, the coefficients $\bm{w}$ are optimized based on a convex combination of energy and forces errors (Eq.~\ref{eq:loss_function}), whose gradient can be computed analytically. This, together with the convexity of RF models, allowed us to write a closed-form expression that {\em globally} minimizes the error without the need to perform gradient descent.
Furthermore, the evaluation of the minimizer can be easily parallelized over multiple GPUs, resulting in virtually linear scaling (\cref{fig:franken-diagram}B).
The resulting model can then be easily deployed for molecular dynamics using the Atomistic Simulations Environment (ASE)~\cite{ase-paper} suite and also the large-scale atomic/molecular massively parallel simulator (LAMMPS)~\cite{LAMMPS} MD engine. Since it is not necessary to perform a full forward pass through the GNN backbone but only up to the layers used for constructing the representation, \algo~models attain higher inference rates compared to the full backbone GNN model (\cref{fig:franken-diagram}C).

In the following, we focus mostly on backbones obtained from MACE-based potentials~\cite{batatia2022mace}, and particularly the MACE-MP0~\cite{batatia2023foundation} invariant model ($L=0$). This is a general-purpose potential trained on MPTraj~\cite{Chen_2022}, containing extended (periodic) systems drawn from the Materials Project,~\cite{materials_project} calculated at the DFT level using the Perdew–Burke–Ernzerhof (PBE)~\cite{PBE_og} exchange and correlation functional.  The training set covers 89 elements of the periodic table, which makes (at least in principle) its use possible in many different applications, from materials to aqueous systems and from interfaces to catalytic systems. In fact, in order to use it as a backbone, the potential must have been trained on some configuration that contained the chemical species of interest.
In the SI, we also compare with equivariant models ($L=1$), other GNN architectures (SevenNet0), and training datasets (MACE-OFF).

\begin{figure*}[t!]  
    \centering
    \includegraphics[width=0.9\textwidth]{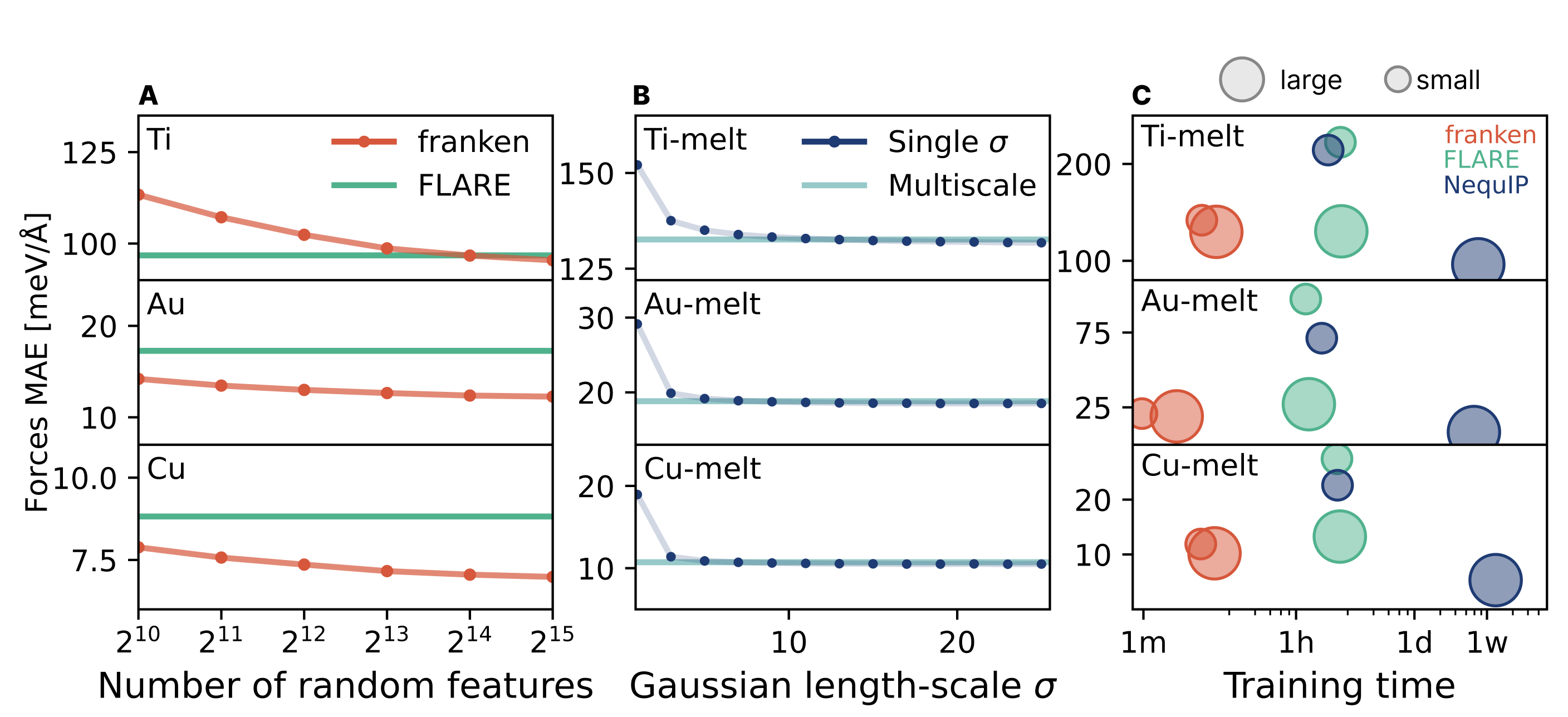}
    \caption{\paragraphtitle{Easy hyperparameter tuning}, shown using selected elements (Ti, Au, Cu) of the TM23 dataset. Panel (A) shows how performance improves monotonically with increasing random features. In panel (B) we validate the effectiveness of multiscale random features that eliminate the need for precise length-scale selection. Panel (C) illustrates the comparison between force predictions and training times vs model complexity for three different systems (Ti, Au, Cu), for which more detailed studies were reported in Ref.~\citenum{owen2024complexity}. For each model, we report results by increasing the complexity of the model (denoted by the size of the circles). For NequIP, this correspond to changing the tensor rank from invariant features ($L=0$) to equivariant ones  ($L=3$ for Ti, Cu and $L=5$ for Au); for FLARE, these are 2-body (B1) and 3-body (B2) representations; for \algo, we changed the number of random features from 512 to 32k.} 
    \label{fig:TM23-elements}
\end{figure*}

\vspace{-1em}
\subsection*{TM23 dataset -- speed and accuracy across the periodic table.}
\vspace{-1em}
To demonstrate the soundness of our approach, we applied it to every one of the 27 transition metals from the TM23 dataset~\cite{owen2024complexity}, comparing both training time and force prediction accuracy against the reported baselines. These are (i) FLARE~\cite{vandermause2020fly}, a kernel method based on ACE descriptors, and (ii) NequIP~\cite{batzner2023nequip}, an equivariant GNN architecture. The Authors of the TM23 dataset poured considerable effort into training the baselines and performed extensive hyperparameter tuning to get the most accurate models possible, oftentimes sacrificing training speed for accuracy (see~\cref{fig:TM23-elements}c). In the following, we report results obtained using the MACE-MP0 ($L=0$)~\cite{batatia2023foundation} model as the backbone block. In the SI, we also report results obtained with the SevenNet-0~\cite{park2024scalable} backbone (\cref{fig:SI-TM23-sevennet}), a model based on the NequIP architecture which has been trained on a Materials Project-based dataset.

In ~\cref{fig:TM23-efficient}, we report the training times (A), the accuracy on forces (B), and energies (C) for each one of the transition metals in the TM23 dataset. While the training times for \algo~ are in the ballpark of a few minutes, including full hyperparameter tuning, FLARE's and NequIP's training times do not include hyperparameter tuning, and extend from a range of a few hours up to weeks. Furthermore, the reported NequIP training times are conservatively estimated according to the reported training procedure, making these numbers realistic in case the models converged within the first epoch of training. These considerations project the cumulative training time of FLARE and NequIP models in weeks if not months. 

In terms of energy/forces prediction, the NequIP baseline is unquestionably more accurate than both FLARE and \algo. As discussed below and showed in~\cref{fig:TM23-efficient}, however, NequIP's accuracy is highly correlated with the complexity (that is, the angular resolution $L$) of the model, and rapidly declines by taking more moderate, and arguably practical values of $L$, see~\cref{fig:TM23-elements}C. FLARE and \algo, on the other hand, attain similar accuracy on the forces with \algo~showing an edge in 24 out of the 27 elements and force errors which are on average $\sim$10\% lower, see the detailed periodic table reported in~\cref{fig:TM23-efficient}D. It is important to note that the FLARE results were obtained by searching, for each system, the set of ACE descriptors that more faithfully represent the local environments. This resulted in heterogeneous models with degrees of expansion of the radial basis ranging from 7 to 17, from 3 to 6 for the angular one, and with a cutoff from 4 to 7 \AA. In contrast, our transfer learning approach requires no choice of system-dependent parameters, providing a general representation that is as good as the best representation based on physical descriptors. This is a considerable advantage to traditional methods, especially in the case of multi-component systems, where it is even harder to search for the optimal parameters. 

On the other hand, \algo's models have very few hyperparameters, considerably reducing the overall cost of training a production-ready MLIP. The "size" of \algo~models is solely determined by the total number of random features $D$, a parameter that can be generally set to be as large as possible (within memory constraints), as we observed a monotonic increase in accuracy as a function of $D$, see \cref{fig:TM23-elements}A. The only other key parameter that needs to be tuned for optimal performance is the length scale of the kernel $\sigma$. As $\sigma$ influences the random features $\phi_{n}(\bm{R})$, which are the most expensive quantities to evaluate at training time, we came up with a {\em multiscale} approach (Eq.~\ref{eq:multiscale}) to avoid costly grid searches over $\sigma$ while keeping the impact on final performances minimal, see \cref{fig:TM23-elements}B. In a nutshell, in the multiscale approach, we select a handful of values of $\sigma$ and assign each of them to a portion of the total random features, see the Methods section for further details. 

We conclude our analysis by highlighting that the performance of both NequIP and FLARE changes quite dramatically with the model complexity, unlike \algo, which remains stable. For example, using smaller versions of the NequIP architecture to speed up the training time can seriously impact the final force prediction error. This is clearly shown in~\cref{fig:TM23-elements}C, where we study the trade-off between training time and force accuracy upon changing the model complexity. For \algo, we investigated the effect of raising the number of random features from 512 up to 32k. For FLARE, we changed the body order of the ACE descriptors from 1 to 2, while for NequIP, we report the results from Ref.~\citenum{owen2024complexity} upon changing the angular resolution $L$. While \algo~is fairly insensitive with respect to the training time and final force accuracy, we see that the force MAE of NequIP doubles by going to $L = 0$ with training times still on the order of one day. In summary, all these results show that \algo~allows the MLIPs to be trained extremely efficiently while providing performances that are as accurate as that of the best kernels built on physical descriptors but without the unfavorable scaling and limitations or the need to manually tune the hyperparameters of the model.

\vspace{-1em}
\subsection*{Bulk Water -- Data efficient potentials for molecular dynamics} 
\vspace{-1em}
\begin{figure*}[t]  
    \centering
    \includegraphics[width=0.95\textwidth]{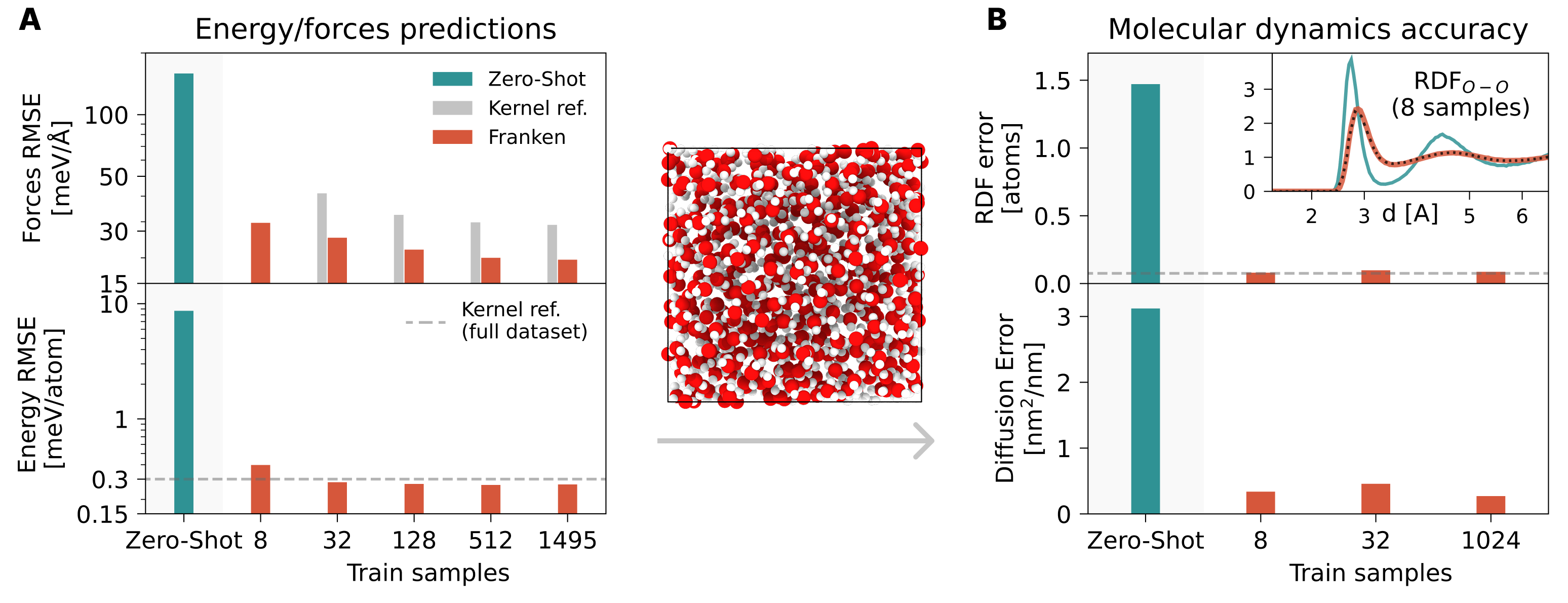}
    \caption{\paragraphtitle{Data efficiency on bulk water}. Left: Root mean squared error (RMSE) in force (top) and energy (bottom) predictions for MACE-MP0 ($L=0$) zero-shot model,  \algo~with MACE-MP0 ($L=0$) backbone, and the reference kernel optimized in Ref.~\citenum{montero2024comparing}. \algo~shows rapid improvement with only a few training samples, outperforming the kernel baseline trained on the full dataset. Center: a snapshot of water MD simulation produced with \algo, from which structural and dynamical properties are computed. Right: accuracy of molecular dynamics simulations assessed via the error on the radial distribution function (RDF, top) and diffusion coefficient (bottom). The RDF error is evaluated against the DFT data, while the diffusion from the results obtained with the kernel in Ref.~\citenum{montero2024comparing}. Inset: oxygen-oxygen RDF from simulations with MACE-MP0 zero-shot and \algo~trained on 8 samples. The black dotted line represents the DFT reference line as reported in Ref.~\citenum{montero2024comparing}. \algo~delivers stable and accurate RDFs with as few as 8 configurations, demonstrating strong out-of-distribution generalization and correction of the MACE-MP0 overstructured water prediction.}
    \label{fig:bulk-water}
\end{figure*}
Besides showing the accuracy and efficiency of ~\algo~in predicting energies and forces, the real benchmark for an MLIP is the ability to run stable and physically meaningful MD simulations. In fact, a low error on force predictions is not enough to judge simulation stability or thermodynamic accuracy~\cite{fu2023forces}. To test \algo's performance on molecular dynamics, we started with bulk water. The motivation for choosing this system is twofold. On the one hand, it serves to test the performance of the transfer learning approach outside the training domain, as it is out of distribution with respect to the systems (inorganic crystals) used to train the MACE-MP0 backbone, unlike the TM23 dataset tested previously. On the other hand, water is highly sensitive to the details of the electronic structure used. Indeed, the PBE exchange-correlation functional, which is used to optimize the MACE-MP0 model, yields an overstructured water with a tendency to predict a stronger hydrogen bonding network than experimentally observed~\cite{lin2012structure}. This leads to the radial distribution function (RDF) peaks being higher than the experimental data and to a slowed mobility, with a significantly underestimated diffusion coefficient. 

We therefore explored the ability of \algo~to transfer MACE-based representations to a dataset of DFT calculations made with a level of theory that describes water more closely to the experiments. To this extent, we used the dataset curated by Montero \textit{et al.}\cite{montero2024comparing}, which is characterized by high-quality DFT calculations (performed using RPBE + D3) and a thorough sampling. Here, the Authors compared the performances of Behler-Parrinello neural networks and kernel-based MLIPs built from physical descriptors, concluding that they both provided practically indistinguishable results, although the NN-based models trained on smaller datasets were unstable in some simulations. For this reason, in the following, we will compare with the kernel results.

\begin{figure*}[t!]  
    \centering
    \includegraphics[width=0.95\linewidth]{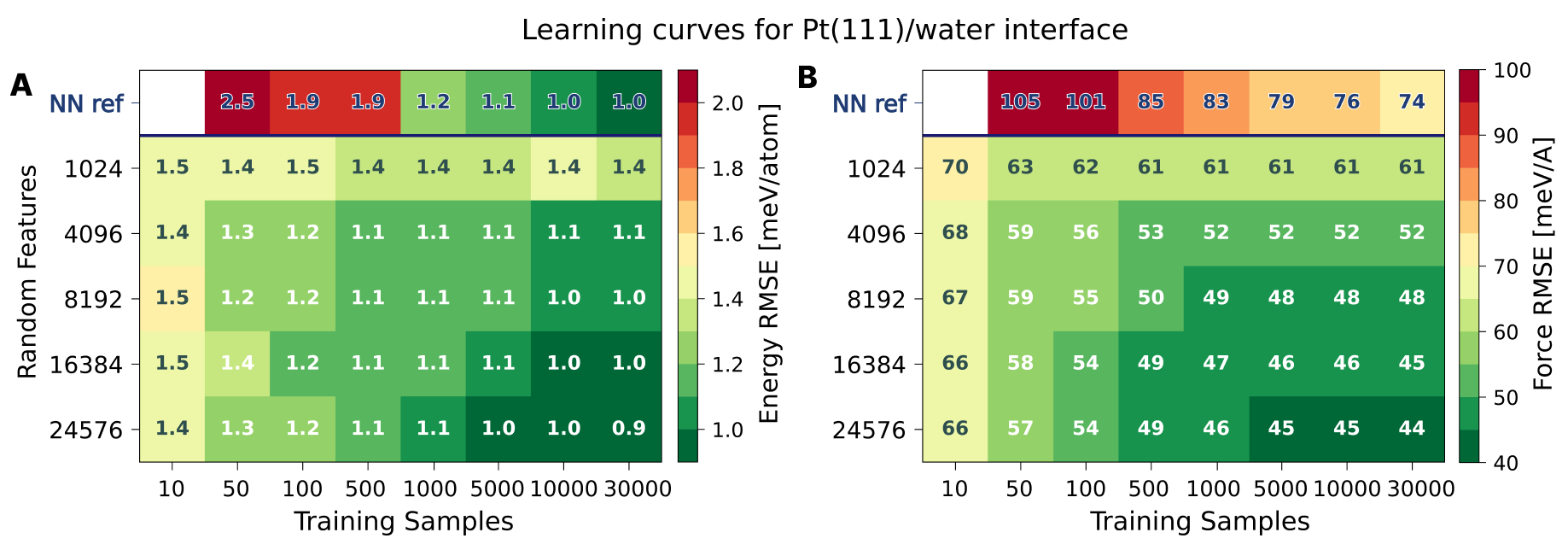}
    \caption{\paragraphtitle{Data efficiency for the Pt(111)/water interface}.
Learning curves for (A) energy and (B) force predictions as a function of number of samples (x-axis) and number of random features (y-axis) used for the \algo~model using the MACE-MP0 backbone. The first line contains the results reported in Ref.~\citenum{mikkelsen2021water}
using the Behler-Parrinello neural network potential (which starts from 50 samples). Each cell is colored according to the RMSE on the energy and forces whose number is reported within the cell. \label{fig:interface-1}
}
\end{figure*}

For this system, our main goal is to assess the data efficiency of \algo~MLIPs. To this effect, we trained the \algo~models on different random subsets of the training data, containing as little as 8 samples, and calculated the root mean square error (RMSE) on the test set, see the left panel of~\cref{fig:bulk-water}. For completeness, we also report the zero-shot predictions, where the energy is calculated with respect to a common reference. Comparing the errors on energy and forces against the results reported for the kernel method in Ref.~\citenum{montero2024comparing}, we observe that with as little as 8 training samples \algo~is able to obtain a high accuracy (forces RMSE equal to 30 meV/\AA), comparable to what was obtained by the reference kernel using {\em the whole training dataset} (27 meV/\AA). Similarly, in the case of energy predictions, with 8 samples \algo~attains an RMSE of 0.4 meV/atom, surpassing the kernel baseline trained on the full training dataset (0.3 meV/atom) already with 32 samples. Energy and forces RMSEs monotonically decrease by training \algo's models on larger training sets. Comparing the learning curves of the force predictions as a function of the number of samples (top left panel of \cref{fig:bulk-water}), we observe that \algo~is more data efficient than training the reference kernel models from scratch.

A competitive accuracy on energy and forces, however, is only part of the story, as we are mostly interested in the ability to produce stable and accurate molecular dynamics simulations. 
For this reason, we ran multiple 1~ns-long MD simulations for each model, which we used to evaluate static (radial distribution function) and dynamic (diffusion coefficient) properties. The first result is that all simulations, even those trained on 8 samples, are stable throughout their duration. On the right side of~\cref{fig:bulk-water}, we show how these MD-related observables change as a function of the number of training samples. It is remarkable that already with only 8 samples, \algo~is able to match the results reported in Ref.~\citenum{montero2024comparing} even for molecular dynamics, obtaining reported results indistinguishable from the reference.  All this despite the fact that the zero-shot description of both properties was significantly off, overestimating the radial function and underestimating the atomic diffusivity. Nevertheless, with only a handful of samples, our transfer learning approach achieved an excellent result. In the SI, we report the full partial radial distribution function (~\cref{fig:SI-water-rdf}), contrasted with the DFT reference data and the experimental data.

The data shown in \cref{fig:bulk-water} is relative to \algo~with MACE-MP0 backbone at $L = 0$. In the SI, we extend our analysis to different GNN backbones, both invariant ($L=0$) and equivariant ($L = 1$), and using MACE-MP0 as well as MACE-OFF~\cite{kovacs2023mace} pre-trained models. The latter were trained on the SPICE~\cite{Eastman_2023_spicev1} dataset of organic molecules and also water clusters. Studying \algo's learning curves as a function of the number of random features and number of samples, we show that they are fairly independent from the chosen backbone, reflecting the generality of our transfer learning approach (\cref{fig:SI-water-backbones}). In the case of MACE-MP0, the gain in accuracy in using equivariant models to predict energy and forces is marginal, while it is more significant in the case of MACE-OFF. In Ref.~\citenum{kovacs2023mace}, the Authors noted that although the MACE-OFF models were trained on small clusters of water to improve the accuracy of solvated systems, the description of water for the MACE-OFF-small invariant model was not excellent, and larger equivariant models had to be used to obtain a good description. In terms of molecular dynamics results, however, when using the MACE-OFF-small as the backbone of \algo~we are able to cure these shortcomings with a handful of samples, obtaining a result indistinguishable from that obtained with the heftier MACE-MP0 (\cref{fig:SI-water-sample-efficiency}). Our transfer learning approach thus allows the use of cheaper (that is, invariant) architectures, which are more computationally efficient in inference without sacrificing accuracy; see \cref{table:SI-timings} for the detailed timings.

\vspace{-1em}
\subsection*{Modeling water-metal interfaces: the case of Pt(111)}
\vspace{-1em}
To further increase the complexity of our benchmark, we extend the evaluation of our approach from isolated transition metals and bulk aqueous systems to the more challenging case of water–metal interfaces. These interfaces play a pivotal role in a broad range of scientific and technological domains, including (electro)catalysis, batteries, and corrosion processes.~\cite{hodgson2009water,nagata2016molecular} However, accurately capturing their behavior remains a significant challenge due to the complex interplay between water structuring, surface chemistry, and long-timescale dynamics. 
To investigate this regime, we selected the Pt(111)/water interface as a representative test case, using the dataset developed in Ref.~\citenum{mikkelsen2021water}. Using Behler-Parrinello neural network potentials, the Authors have reported that the Pt(111)/water interface features a double adsorption layer, with a tightly bound primary layer and a more weakly bound one. The primary layer is anchored by strong Pt–O interactions, while hydrogen bonds link it to the secondary one, and an effective repulsion between adsorbed molecules creates a semi-ordered structure. To achieve these results, they created a training set of 50,000 structures obtained through an active learning process using the NN potential and MD simulations. The RMSE errors of the network in predicting energies and forces are $\sim1$ meV/atom and $\sim70$ meV/\AA, respectively. Interestingly, the Authors reported that training with a few hundred structures was not sufficient to generate stable molecular dynamics simulations and that they had to use the entire data set.

As in the previous example, we evaluated the sample efficiency of \algo's approach, starting with an assessment of the model’s accuracy in reproducing energies and forces using the MACE-MP0 invariant model. In \cref{fig:interface-1} we report the learning behavior as a function of the number of training samples and the number of random features. It is worth observing that energies are very rapidly learned and are rather insensitive to the number of random features, with RMSE values immediately converging to  $\sim$1 meV/atom. In the case of forces, there are more considerations to be made. On the one hand, we observe high accuracy and data efficiency, which leads already with 10 samples to a lower  RMSE than the one obtained from the reference over the whole dataset. 
On the other we see a monotonic improvement as the number of random features increases, saturating around $\sim45$ meV/\AA~ with 16k RFs. We also observe how the number of samples needed to converge accuracy increases - rightly so - with the number of parameters to be optimized: using 1024 RFs already with 50 points one achieves a result within 2 mev/\AA~from that over the whole dataset (63 vs 61 meV/\AA), while with 24k RFs one needs to go up to 1000 points to get a similar result (46 vs 44 meV/\AA), but which at the same time is also considerably more accurate in absolute terms. 

\begin{figure}[t!]  
    \centering
    \includegraphics[width=0.9\linewidth]{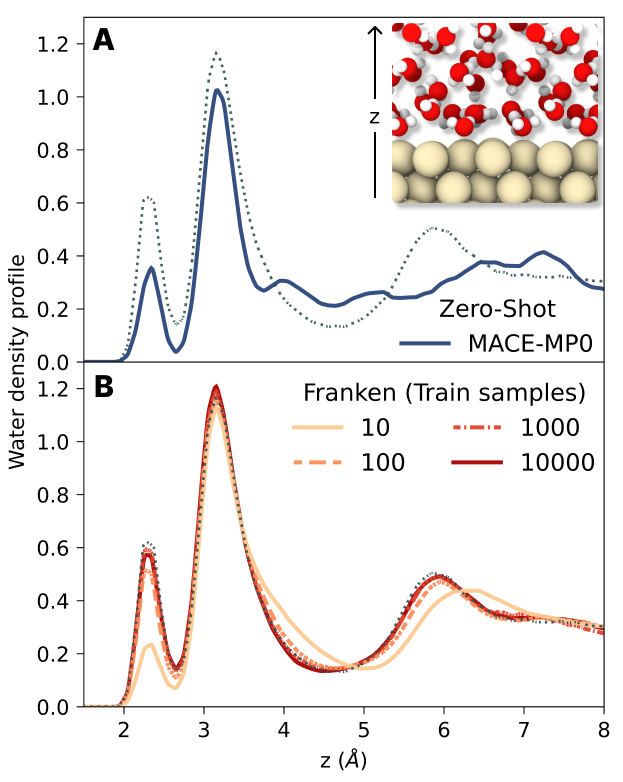}
    \caption{\paragraphtitle{Structural characterization of the Pt(111)/water interface}.
(A): Oxygen density profile along the $z$ direction normal to the Pt(111) surface, comparing predictions from a zero-shot MACE-MP0 model (solid line) to a reference obtained with an independent model trained on the full dataset (dotted line). The MACE-MP0 model captures basic features but deviates in peak heights and fine structure.
(B): Density profiles predicted by \algo~models trained using 24k RFs on an increasing number of samples (solid lines, from light to dark colors); with as few as 100 samples, these models accurately match the reference. In the inset, we report a snapshot from a molecular dynamics simulation showing the structure of the two layers composing the interface, the first with a strong chemisorption and the second with a weak physisorption.}
    \label{fig:interface-2}
\end{figure}

More importantly, we assessed its performance in capturing structural properties via molecular dynamics. In the top panel of~\cref{fig:interface-2}, we report the density profile of oxygen along the direction perpendicular to the surface as generated with MACE-MP0 zero-shot. This profile has some correct qualitative features (such as the presence and the location of the two peaks) but lacks many quantitative ones (such as the height of the peaks and the structure of water at medium range). In the case of~\algo, already when trained with 10 samples, it captures most of the features of the density profile, while from 100 onwards, it correctly reproduces the reference data, including the characteristic density layering of water near the platinum surface into two distinct peaks, featuring strong and weak adsorption respectively.
The results again underscore the exceptional data efficiency of our transfer learning approach, not only in learning energy and forces but also in producing stable molecular dynamics simulations and accurate physical observables from a minimal dataset. Hence, it paves the way for the modeling of more complex and realistic interfaces, which require the inclusion of many different ingredients, from interactions with adsorbates to including defects and ions in water, which are key aspects of many relevant applications.

\vspace{-1em}
\section*{Discussion}\label{sec:discussion}

In this work, we introduced \algo, an efficient and scalable transfer learning framework for machine learning interatomic potentials. By leveraging invariant representations learned from pre-trained graph neural networks combined with random Fourier features, \algo~delivers an algorithm that is remarkably fast to train, easy to optimize, and highly scalable across multiple GPUs. 

The reasons for the success of this approach are twofold. On the one hand, we exploit the ability of GNNs to learn in a data-driven way an effective representation of local environments. On the other hand, the random features approach allows this representation to be transferred to new systems in an extremely efficient way. By effectively reusing information that has already been learned, we need only a few training samples to adapt it to new systems. Of course, these two aspects are closely related, since without a good representation, the transfer learning approach would be less effective. While we have focused mainly on the MACE architecture in this manuscript, it will be interesting to understand the impact of other architectures on the resulting latent representation in the future. Similarly, the availability of models trained on larger datasets will also allow us to study the dependence on the number and type of data.

The data-efficiency of \algo~opens up a variety of practical applications. One notable area is the straightforward transfer of general-purpose potentials to higher levels of quantum mechanical theory or systems outside their original training scope. Our results on bulk water demonstrate the high efficiency in adapting the potential to a different DFT functional, and could be successfully employed to bring higher accurate methods beyond DFT on complex systems~\cite{chen2023data,bocus2025operando}. Another promising application is realistic modeling of complex catalytic systems or materials featuring numerous distinct chemical environments where \algo's computational efficiency and adaptability become particularly valuable. Even more, the sample and computational efficiency make it ideally suited for high-throughput screening workflows, enabling rapid and accurate prediction of properties across large libraries of candidate materials or catalysts. Finally, although this work focused on materials and condensed matter systems, \algo's features—such as its independence from the number of chemical species and high training efficiency—make it suitable for biomolecular systems as well. With an appropriate backbone, it can be extended to systems such as small peptides, proteins, and metalloenzymes, offering a promising route toward data-efficient force fields for biochemistry and drug discovery.

Moving forward, several developments could enhance \algo's capabilities and broaden its applications. For more complex processes such as chemical reactions, where most of the information lies in the tails of the distribution, sampling and selecting the relevant structures becomes as crucial as having data-efficient architectures.~\cite{perego2024data} 
Combining it with active learning strategies would enable efficient exploration of relevant configurations, in turn facilitating accurate modeling of rare events such as chemical reactions, phase transitions, and complex dynamical processes. 
Another interesting direction is exploring alternative pre-trained backbones derived from self-supervised or meta-learning techniques \cite{hospedales2021,falk2022}. Such methods could yield more convenient representations, expanding the scope of applicability to different and more complex physical and chemical systems. 

Finally, extending \algo's framework beyond the prediction of energies and forces represents an exciting frontier~\cite{sipka2023constructing,pengmei2025using}. Applying our transfer learning approach to chemical properties prediction or reaction coordinate identification could provide substantial advancements in computational chemistry and materials science. By adapting pre-trained representations to these distinct but related predictive tasks, \algo~could become a versatile tool, bridging the gap between accurate quantum mechanical calculations and practical, large-scale molecular simulations.

\section*{Methods}\label{sec:methods}

\subsection*{Extracting GNN representations}
\vspace{-1em}
\noindent\algo~is able to transfer knowledge from large pre-trained GNN MLIPs to new physical systems by extracting invariant features from the inner layers of these pre-trained models. As GNN architectures usually consist of multiple hidden layers concatenated to each other, a basic strategy to get a GNN representation is to stop the forward pass to the chosen layer $\ell$, and take the {\em invariant} node features for that layer as a representation. This is the approach already introduced in~\cite{falk2024transfer}, which we followed for the SevenNet-0 on the TM23 dataset. In the case of backbones from the MACE family~\cite{batatia2022mace}, instead, we extracted a representation consisting of the concatenation of the node features {\em up to the chosen layer} $\ell$. This strategy mimics the structure of the MACE architecture itself, where the node features from every hidden layer are first concatenated and then fed to a readout layer for atomic energy prediction. Although we focus on invariant descriptors, in cases where the GNN backbone employs equivariant message-passing schemes~\cite{batatia2022mace, batzner2023nequip}, the invariant features of the inner layers are determined by the underlying equivariant information from previous layers. This enables \algo~to leverage equivariant properties indirectly while maintaining the computational simplicity of invariant features.

Before being fed to the RF block (see the second panel of~\cref{fig:franken-diagram}), we standardize the GNN features by subtracting their mean and scaling them by their standard deviation. In systems with multiple chemical elements, we found that {\em per-species} standardization works particularly well -- here, the mean and standard deviation used to scale the features of an atom of species $s$ are computed only over atoms of the same element $s$. This standardization process is quite important in practice, as RF models work best when their inputs are properly centered and scaled.
\vspace{-1em}

\subsection*{Random Features}
\vspace{-1em}
Random Fourier Features offer a framework to approximate kernel functions $k(h, h')$ by expressing them as scalar products in a finite-dimensional feature space:
\begin{equation}
    k(h, h') \approx \phi(h)^\top \phi(h').
\end{equation}

To compute such feature map, \citet{rahimi2007random} proposed to leverage the Bochner theorem~\cite{rudin2017fourier}, which states that any continuous, stationary kernel can be represented as the Fourier transform of a non-negative measure. Hence, we can construct the kernel starting from its Fourier transform. For instance, the Gaussian kernel,
\begin{equation}
k(h, h') = \exp\left(-\frac{\|h - h'\|^2}{2\sigma^2}\right),
\label{eq:gaussian}
\end{equation}
has a Fourier transform that is itself Gaussian, and so it can be approximated by averaging cosine functions with frequencies sampled from a standard normal distribution. In practice, this leads to a randomized feature map of the form~\cite{rahimi2007random}:
\begin{equation}
\phi(h) = \sqrt{2} \cos\left(\frac{\bm{W} \cdot h}{\sigma} + \bm{b}\right),
    \label{eq:rfs}
\end{equation}
where each row of \( \bm{W} \in \mathbb{R}^{D \times d} \) is drawn from a standard multivariate normal distribution, and the offset vector \( \bm{b} \in \mathbb{R}^D \) has components sampled uniformly from \( [0, 2\pi) \)~\cite{rahimi2007random,yu2016orthogonal}.  Here, \( D \) is the number of random features, which defines the model complexity and controls the quality of the kernel approximation. In this work, we used orthogonal random features~\cite{yu2016orthogonal}, which yield improved approximation of the Gaussian kernel, by slightly modifying the sampling scheme for $\bm{W}$. Importantly, \( \bm{W} \) and \( \bm{b} \) are just sampled at initialization and do not require training, making the method simple and efficient to implement.

Two observations are now in order. First, replacing the exact kernel with its finite-dimensional approximation $ \phi(h)^\top \phi(h') $ enables learning using only \( D \times D \) matrices. This avoids the quadratic or quartic scaling associated with exact kernel methods, especially in the presence of gradients. For example, computing the whole kernel matrix for both energy and force labels requires evaluating up to \( 9(NT)^4 \) elements where $N$ is the number of atoms and $T$ the number of atomistic configurations in the dataset. In contrast, the RF-based model scales linearly with the number of training points and enables constant-time inference, making it practical even for large datasets and systems. Second, the random feature map can be interpreted as a fixed, one-layer neural network with a cosine activation, applied to the GNN-derived descriptors \( h_n(\bm{R}) \). This observation allows for seamless integration into existing deep learning frameworks.
\vspace{-1em}

\subsection*{Multiscale Random Features}
\vspace{-1em}
In kernel-based models such as \algo, the most critical hyperparameter is the kernel length scale \( \sigma \), which governs the locality of the Gaussian kernel and directly influences the resulting random features \( \phi_n(\bm{R}) \). Since these features are the most computationally expensive part of the model to evaluate during training, tuning \( \sigma \) via grid search can be costly and inefficient.
To address this, we introduced \emph{multiscale random features}, a simple yet effective strategy that allows the model to operate over a range of length scales within a single representation. Rather than relying on a single, optimal \( \sigma \), we construct random features that span multiple scales, thereby eliminating the need for explicit hyperparameter tuning.
Specifically, given a total budget of \( D \) random features, we define a set of \( d \) length scales \( \{ \sigma_1, \sigma_2, \dots, \sigma_d \} \), sampled uniformly within a predefined range. For each \( \sigma_j \), we allocate \( D/d \) features by sampling a corresponding frequency matrix \( \bm{W}_j \sim \mathcal{N}(0, \sigma_j^{-2} \bm{I}) \) where we have absorbed the length scale $\sigma_j$ inside $\bm{W}$ itself.
The final feature map is constructed by concatenating the individual components:
\begin{equation}
    \phi(h) = \left[ \, \phi^{(1)}(h) \; \| \; \phi^{(2)}(h) \; \| \; \dots \; \| \; \phi^{(d)}(h) \, \right],
    \label{eq:multiscale}
\end{equation}
where each block is given by:
\begin{equation}
\phi^{(j)}(h) = \sqrt{2} \cos\left( \bm{W}_j \cdot h + \bm{b}_j \right).
\end{equation}
Empirically, this multiscale approach achieves comparable performance to optimally tuned single-scale models, as shown in panel (B) of~\cref{fig:TM23-elements}, while significantly reducing the tuning effort.
\vspace{-1em}

\subsection*{Model optimization}\label{sec:theory}
\vspace{-1em}
To train \algo~we minimize a convex combination of  $\ell_{\text{E}}$ and $\ell_{\text{F}}$, the standard squared errors on energy and forces, together with a $\text{L}_2$ regularization term on the norm of the weights $\|\bm{w}\|$
\begin{equation}
    \ell_{\alpha}(\bm{w}) = (1 - \alpha)\ell_{\text{E}}(\bm{w}) + \alpha\  \ell_{\text{F}}(\bm{w}) + \lambda\|\bm{w}\|^2.
    \label{eq:loss_function}
\end{equation}
where $\alpha$ is a parameter which determines the relative importance of energy and forces in the loss function. 

The convexity properties of RF models ensure that there exists a vector of coefficients $\bm{w}^{*}$ which {\em globally} minimizes the loss function $\ell_{\alpha}(\bm{w})$. We now compute the global minimizer in the {\em energy-only} case $\alpha = 0$, as the same steps (and slightly more cumbersome calculations) lead to a similar result when forces are present $\alpha > 0$. The energy-only loss over a dataset of reference configurations $\mathcal{D} = (\bm{R}_{i}; E_{i})_{t = 1}^{T}$ is
\begin{equation}
     \ell_{0}(\bm{w}) = \ell_{\text{E}}(\bm{w}) = \sum_{t=1}^{T} \left( E(\bm{R}_{t}; \bm{w}) - E_t \right)^2 + \lambda \|\bm{w}\|^2.
\end{equation}
To minimize $\ell_{\text{E}}(\bm{w})$ let's first recall that for a single configuration $\bm{R}$,
\begin{equation}
    E(\bm{R}; \bm{w}) = \sum_{n=1}^{N}\phi_{n}(\bm{R})^{\top}\bm{w} = \varphi(\bm{R})^{\top}\bm{w},
\end{equation}
where we have defined $\varphi(\bm{R}) = \sum_{n= 1}^{N}\phi_{n}(\bm{R})$. Plugging the definition of the energy back into the loss function and taking the gradient of $\ell_{\text{E}}(\bm{w})$ with respect to $\bm{w}$, one gets:
\begin{align*}
    \label{eq:grad_energy_loss}
     \nabla_{\bm{w}}\ell_{{\rm E}}(\bm{w}) = &2\big[\left(\msf{C} + \lambda \text{I}_{D}\right)\bm{w} - \msf{b}\big],
\end{align*}
where the covariance matrix $\msf{C}$ and coefficient vector $\msf{b}$ are, respectively
\begin{align*}
    \msf{C} &= \sum_{t = 1}^{T}\varphi(\bm{R}_{t}) \varphi(\bm{R}_{t})^{\top} \in \R^{D\times D}, \\ 
    \msf{b} & = \sum_{t = 1}^{T}E_{t}\varphi(\bm{R}_{t}) \in \R^{D}.
\end{align*}
Since the loss function $\ell_{\text{E}}(\bm{w})$ is strongly convex in $\bm{w}$, it has a unique global minimizer corresponding to the solution of  $\nabla_{\bm{w}}\ell_{{\rm E}}(\bm{w}) = \bm{0}$
\begin{equation}\label{eq:energy_minimizer}
    \bm{w}^{*} = (\lambda \text{I}_{D} + \msf{C})^{-1}\msf{b}.
\end{equation}
The linear system in~\cref{eq:energy_minimizer} has a dimension equal to the number of random features $D$, for which a reasonable size is on the order of a few thousand, making its solution extremely fast.  As it turns out, the computational bottleneck in computing $\bm{w}^{*}$ lies in the evaluation of the random features $\phi_n(\bm{R})$ for every configuration in the training dataset. Indeed, when forces are taken into account, the descriptor calculation involves differentiating through the GNN backbone and usually comprises more than 99\% of the total training time. Luckily, this process can be carried over in parallel over multiple GPUs, by assigning a subset of the training points to evaluate to each GPU. This trick speeds up the training process almost linearly with the number of GPUs, see panel (B) of~\cref{fig:franken-diagram}.
\vspace{-1em}

\subsection*{Dataset and training details}
\vspace{-1em}

We report here the training procedure for the \algo's models. Unless otherwise stated, we trained the MLIPs using the invariant MACE-MP0-L0 GNN backbone, using as atomic descriptors the node features concatenated up to the second interaction block. A multiscale Gaussian kernel was employed, using four sets of length scales ranging from 8 to 32. Loss hyperparameters were automatically optimized via efficient grid search (does not requires recalculating the covariance matrix). In particular, we tested force-energy weight ratios $\alpha$ from a list of values [0.01, 0.5, 0.99] and the $\text{L}_2$ regularization $\lambda$ from [$10^{-6}$,...,$10^{-11}$]. In the following we describe the datasets used and further training details. 

\paragraphtitle{TM23.} This dataset, taken from Ref.~\citenum{owen2024complexity}, comprises 27 elemental systems, each obtained from AIMD simulations performed at three different temperatures, going from the crystal phase to the liquid: 0.25 $T_m$, 0.75 $T_m$ and 1.25 $T_m$. From these trajectories, 1000 structures per system were selected and calculated with the PBE exchange and correlation functional (non-spin polarized calculations). Of these, 100 structures were used for validation. The number of atoms ranged from 31 to 80 depending on the chemical system.
For \algo, we used the same training (2700) and validation (300) splits as in the dataset paper. Additional training runs were performed  on the melt (1.25 $T_m$) dataset for Ti, Au, and Cu to align with published benchmarks by changing the model complexity.

\paragraphtitle{Water.} The training dataset was adopted from Ref.~\citenum{montero2024comparing} and comprises 1495 curated structures generated through enhanced sampling simulations and active learning. Energies and forces were calculated with DFT using the RPBE+D3 exchange-correlation functional. Data acquisition involved replica exchange molecular dynamics and Bayesian on-the-fly learning as implemented in VASP,~\cite{kresse1993ab, kresse1994ab,kresse1999ultrasoft, kresse1996efficient, kresse1996efficiency, kresse1994norm} resulting in a diverse set configurations spanning a broad thermodynamic range. In addition to the main training dataset, a second dataset of 189 structures was independently generated using temperature-ramped simulations. To better assess model generalizability, we used the first dataset as  training set and the second one as validation. The number of random features used is equal to 8192, unless otherwise specified for the studies as a function of the number of RFs parameters. 
Sample complexity studies were also performed by training on subsets of the first dataset ranging from 8 to 1495 structures.

\paragraphtitle{Pt/water interface}. The training dataset for the Pt(111)/water interface is generated in Ref.~\citenum{mikkelsen2021water}, comprising 48,041 configurations of 144-atom systems (3×4 Pt(111) slab with 48 Pt atoms and 32 water molecules). These structures were generated iteratively starting from AIMD dataset and using active learning with a neural network-based MLP which was used to generate uncorrelated configurations from a series of 10ns-long MD simulations at T=350 K. Energies and forces were calculated using the PBE functional with D3 van der Waals corrections. For the training of \algo, sample complexity studies were performed by training models on randomly selected subsets ranging from 10 to 30,000 configurations and a validation set of 1000 structures. The effect of model complexity was explored by varying the number of random features, using values of 1024, 8192, 16,384, and 24,576.
\vspace{-1.5em}

\subsection*{Molecular Dynamics settings.}
\vspace{-1em}

\paragraphtitle{Bulk Water.}
Two sets of molecular dynamics (MD) simulations were performed to investigate the structural and dynamical properties of bulk water. For the calculation of radial distribution functions, simulations were carried out in the canonical (NVT) ensemble at a temperature of 325 K, using a Langevin integrator with a friction coefficient of 0.01 fs$^{-1}$. The hydrogen atom mass was set to 2 atomic mass units (amu) and a timestep of 0.5 fs was used. Initial velocities were sampled from a Maxwell-Boltzmann distribution. Each system contained 64 water molecules, and five independent replicas of 100 ps were generated. As shown in Ref.~\citenum{montero2024comparing}), this system size does not exhibit significant finite size effects for RDF calculations. The RDF error is calculated as the square root of the integral of the difference between the calculated RDF and the DFT reference reported in~\cite{montero2024comparing}.
For the evaluation of the diffusion coefficient, larger systems consisting of 504 molecules were used to ensure convergence of the results. These simulations also employed a timestep of 0.5 fs, with hydrogen atoms assigned their physical mass (1 amu). Initial equilibration was performed in the NVT ensemble, followed by production runs in the microcanonical (NVE) ensemble for 200 ps. The absolute error between the calculated coefficient and the one reported for the kernel reference~\cite{montero2024comparing} is reported.

\paragraphtitle{Pt/Water Interface.}
To investigate the behavior of water at a platinum interface due to the longer timescale involved, NVT simulations were carried out for 1 ns at a temperature of 350 K. A timestep of 0.5 fs was employed, with hydrogen atoms having a mass of 2 amu. The temperature was controlled using a Nosé–Hoover thermostat with a coupling constant of 2 ps, and an harmonic restraint was used to avoid the water molecules to diffuse to the other side of the metallic slab. To compute the sample efficiency of MD observables, for each number of random features and number of subsamples we performed 5 MD simulations of 1 ns each. To ensure the same computational setup was used to compare with the reference density profile, we used an independent GNN model trained with MACE on the full training set, since we were not able to produce stable simulations with the NN potential released with the dataset. 
\vspace{-1.5em}

\renewcommand\acknowledgmentsname{Code availability}
\begin{acknowledgments}
\vspace{-1em}
Our implementation of \algo~ is open source and is available at the following link: \href{https://github.com/CSML-IIT-UCL/franken}{https://github.com/CSML-IIT-UCL/franken}. The software can be easily installed via pip (\texttt{pip install franken}). The documentation is available here: \href{https://franken.readthedocs.io/}{https://franken.readthedocs.io/}, which also includes a tutorial that can be run on Google Colab to train the model and deploy it for molecular dynamics.

The implementation is based on PyTorch~\cite{paszke2019pytorch}. To maximize performance and minimize the memory footprint, we wrote custom CUDA kernels for the computation of the covariance, and support multi-GPU training through \texttt{torch.distributed}. Currently, MACE and SevenNet backbones are supported. 
\algo~potentials are interfaced both with the Atomistic Simulation Environment (ASE),~\cite{ase-paper} and, in the case of the MACE backbones, also with LAMMPS~\cite{LAMMPS}.
\end{acknowledgments}
\vspace{-1em}

\renewcommand\acknowledgmentsname{Author contributions}
\begin{acknowledgments}
\vspace{-1em}
L.B. and P.N. designed the project; P.J.B. prepared the prototype of the method; G.M., P.N., P.J.B. and L.B. developed the software and performed the simulations; and all authors analyzed the results. L.B. and P.N. wrote the first draft of the manuscript, and all authors edited it.
\end{acknowledgments}
\vspace{-1em}

\renewcommand\acknowledgmentsname{Acknowledgments}
\begin{acknowledgments}
\vspace{-1em}
We thank Michele Bianchi, Simone Perego, and Enrico Trizio for feedback and discussions. We acknowledge the support of the Data Science and Computation Facility at the Fondazione Istituto Italiano di Tecnologia and the CINECA award under the ISCRA initiative. This work was partially funded by the European Union - NextGenerationEU initiative and the Italian National Recovery and Resilience Plan (PNRR) from the Ministry of University and Research (MUR), under Project PE0000013 CUP J53C22003010006 "Future Artificial Intelligence Research (FAIR)". 
L.R. acknowledges the financial support of the European Research Council (grant SLING 819789), the European Commission (Horizon Europe grant ELIAS 101120237), the Ministry of Education, University and Research (FARE grant ML4IP R205T7J2KP; grant BAC FAIR PE00000013 CUP J33C24000420007 funded by the EU - NGEU). 
\end{acknowledgments}
\vspace{-1em}

\bibliography{biblio} 

\clearpage
\appendix

\renewcommand{\thetable}{S\arabic{table}}
\renewcommand{\thefigure}{S\arabic{figure}}
\renewcommand{\theHfigure}{S\arabic{figure}}
\setcounter{figure}{0}

\onecolumngrid

\section*{Supplementary Material}

\subsection*{TM23}

\begin{figure*}[h!]  
    \centering
    \includegraphics[width=\textwidth]{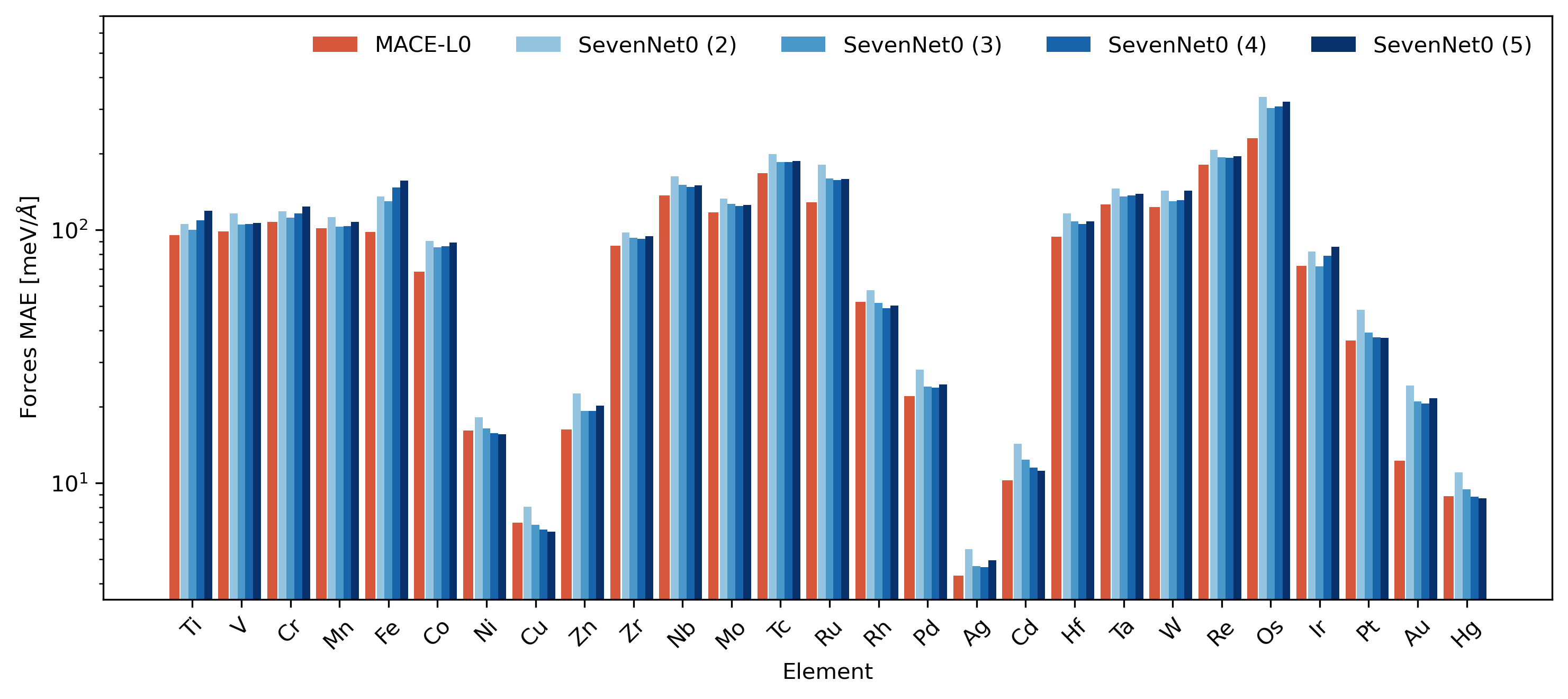}
    \includegraphics[width=\textwidth]{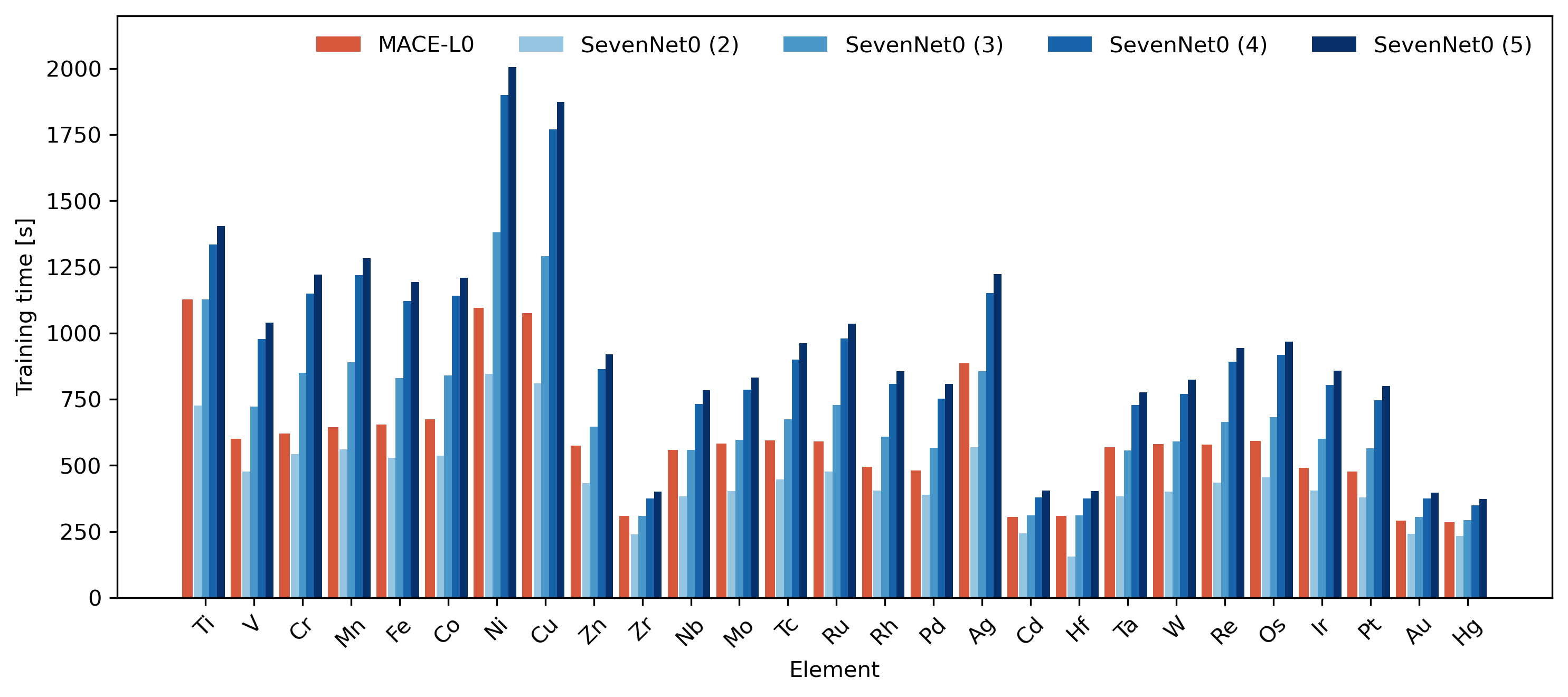}
    \caption{\textbf{Forces accuracy and timings for the SevenNet0 backbone}. (top) Forces MAE for each element of the TM23 dataset.  (bottom) Training times for the SevenNet0 models. In both panels, each bar represent \algo model trained using representations extracted at different interactions layer, from 2 to 5. As a baseline, the MACE-MP0 used in the manuscript is shown in red. Note that for the MACE architecture, we do analyze the dependence on the interaction layer since it only has 2 of them. The behavior along the periodic table between the backbones is very similar, although the SevenNet0-based models are slightly less accurate but also faster if a representation from the first layers is used.}
    \label{fig:SI-TM23-sevennet}
\end{figure*}

\clearpage
\subsection*{Bulk water}
\begin{figure}[h!]  
    \centering
    \includegraphics[width=0.8\textwidth]{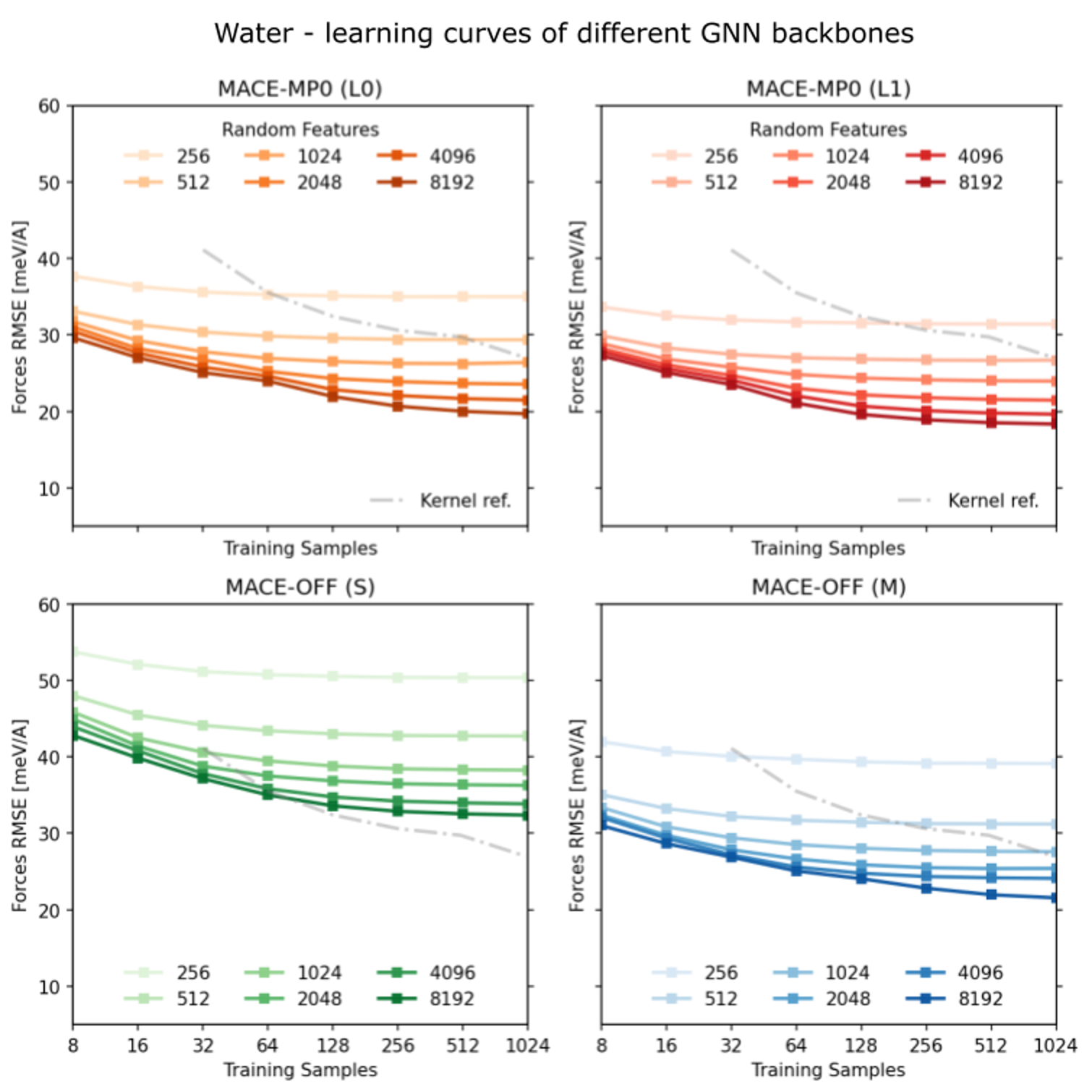}
    \caption{\textbf{Sample complexity studies vs different MACE backbones and number of RFs}. Each panel contains the learning curves in terms of the number of samples for four MACE backbones (MACE-MP0-L0 and L1, MACE-OFF small and medium). Each line correspond to a different number of random features, to show the monotonic increase upon model complexity. }
    \label{fig:SI-water-backbones}
\end{figure}

\begin{figure}[h!]  
    \centering
    \includegraphics[width=\textwidth]{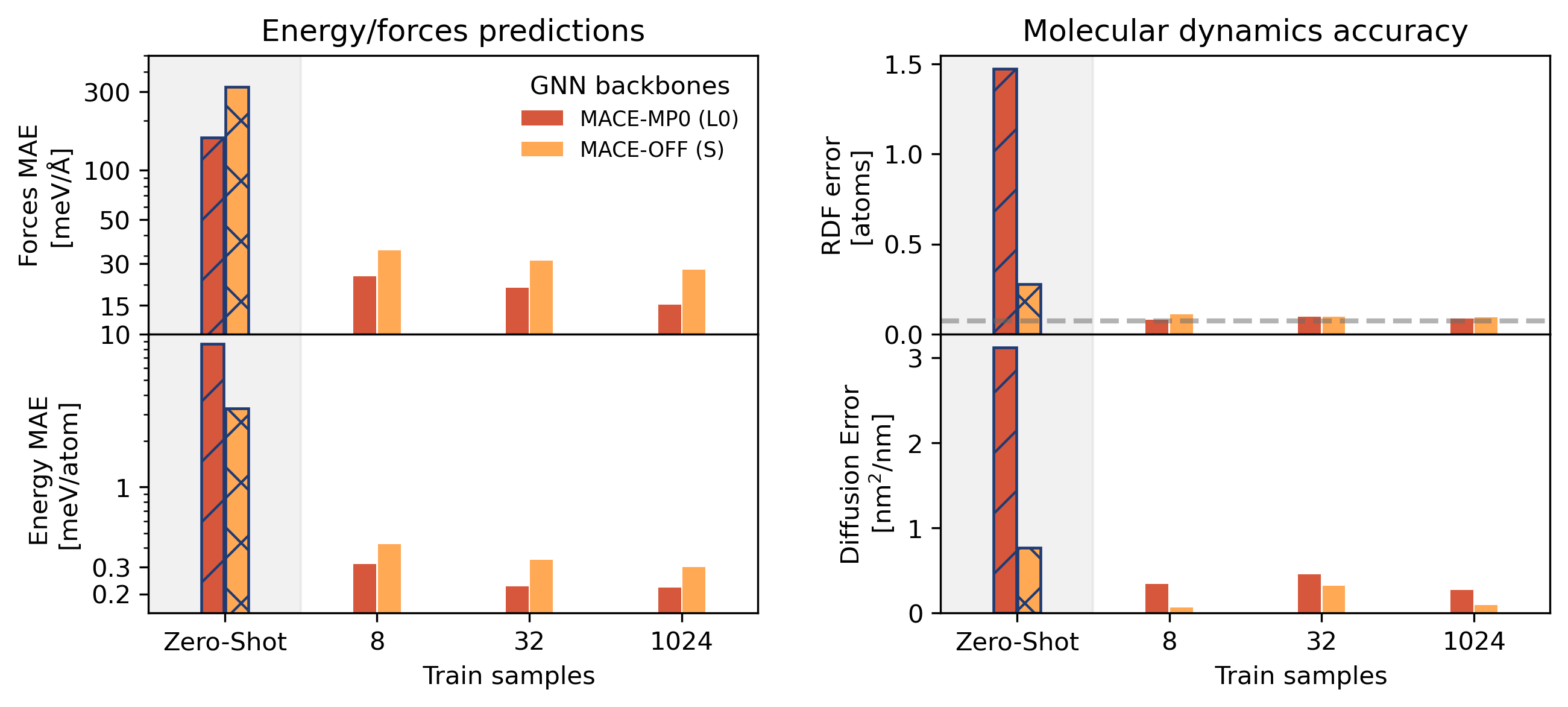}
    \caption{\textbf{Sample efficiency on the water dataset} for force and energy predictions (left) and MD properties (right). Same as~\cref{fig:bulk-water} but for two different backbones: MACE-MP0 L0, red and MACE-OFF-small in orange. The first columns are related to the zero-shot performances.}
    \label{fig:SI-water-sample-efficiency}
\end{figure}

\begin{figure}[h!]  
    \centering
    \includegraphics[width=0.5\textwidth]{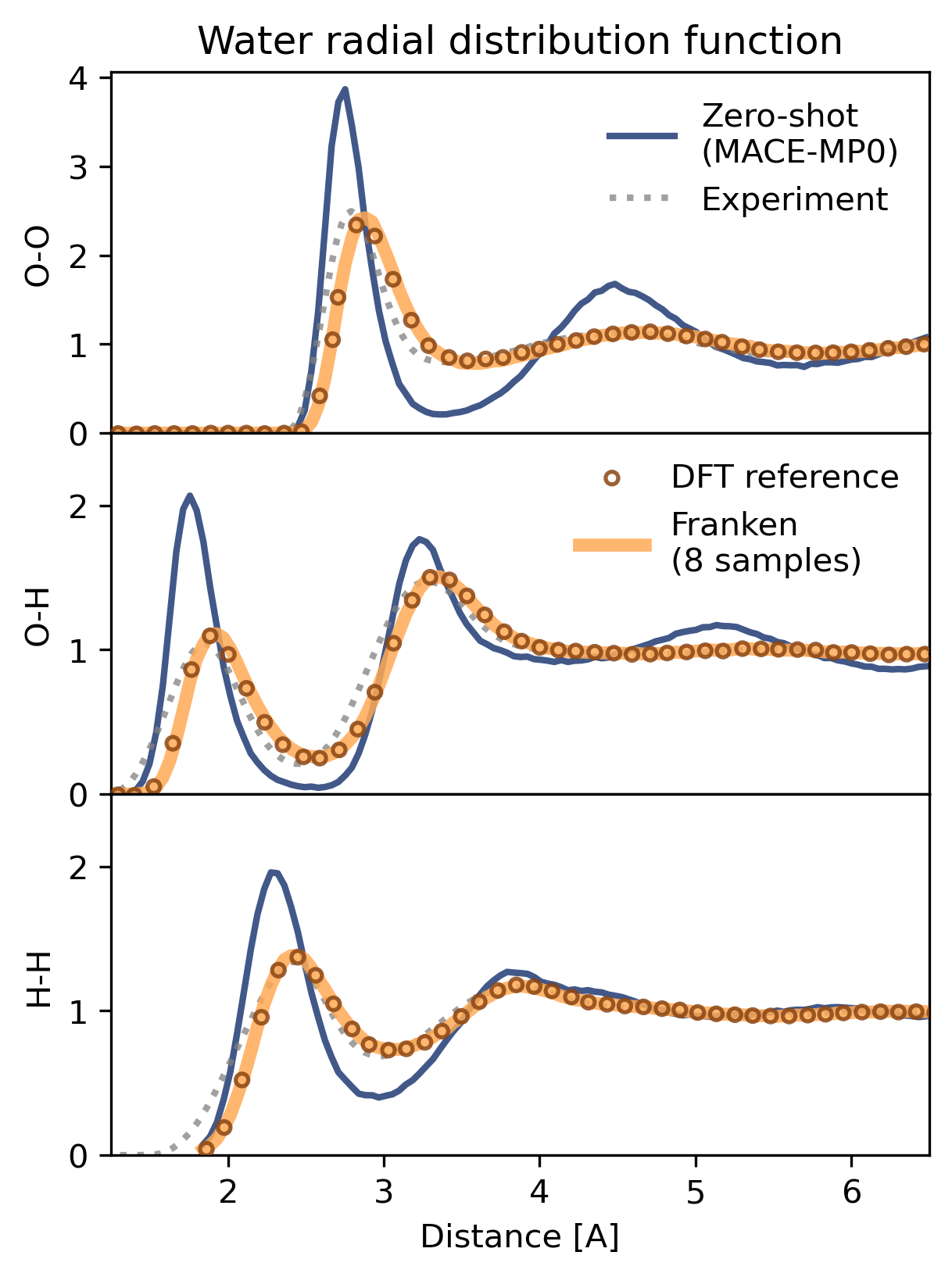}
    \caption{\textbf{Partial radial distribution functions for water} for oxygen-oxygen (top), oxygen-hydrogen (center), and hydrogen-hydrogen (bottom). The solid blue line denotes the zero-shot result obtained with MACE-MP0 $L=0$, while the orange thick line denotes \algo's results using the same backbone, 4096 RFs, and trained on 8 configurations. The empty circles represent the DFT reference, and in a dotted grey line, the experimental curve, both taken from Ref.~\citenum{montero2024comparing}.}
    \label{fig:SI-water-rdf}
\end{figure}

\begin{table}[h!]
\centering
\caption{\textbf{Inference timings for models trained on different MACE backbones}. Comparative inference time benchmark of the different MACE-MP0 (L0 and L1) and MACE-OFF (small and medium) backbones used in this work for zero-shot and ~\algo. Timings were obtained using ASE-based~\cite{ase-paper} calculators such as the \texttt{MACECalculator} as implemented in the MACE~\cite{batatia2022mace} Python package for MACE-based models and our custom \texttt{FrankenCalculator} implementation for \algo ~models predicting energy and forces for a box of water molecules. }
\begin{tabular}{@{}cllll@{}}
\multicolumn{1}{l}{}      &         & \multicolumn{2}{c}{Atoms/sec}    &         \\ \midrule
\multicolumn{1}{l}{Model} &  & Zero-shot & \algo & Speedup \\ \midrule
\multirow{2}{*}{MACE-MP0} & L0      & 8.9      & 11.8                & 1.32x    \\
                          & L1      & 5.4      & 6.4                & 1.18x    \\ \midrule
\multirow{2}{*}{MACE-OFF} & small   & 16.2     & 21.5                & 1.32x    \\
                          & medium  & 7.9      & 9.4                 & 1.2x     \\ \bottomrule
\end{tabular}
\label{table:SI-timings}
\end{table}

\clearpage

\subsection*{Pt(111)/water interface}
\begin{figure}[h!]  
    \centering
    \includegraphics[width=0.45\textwidth]{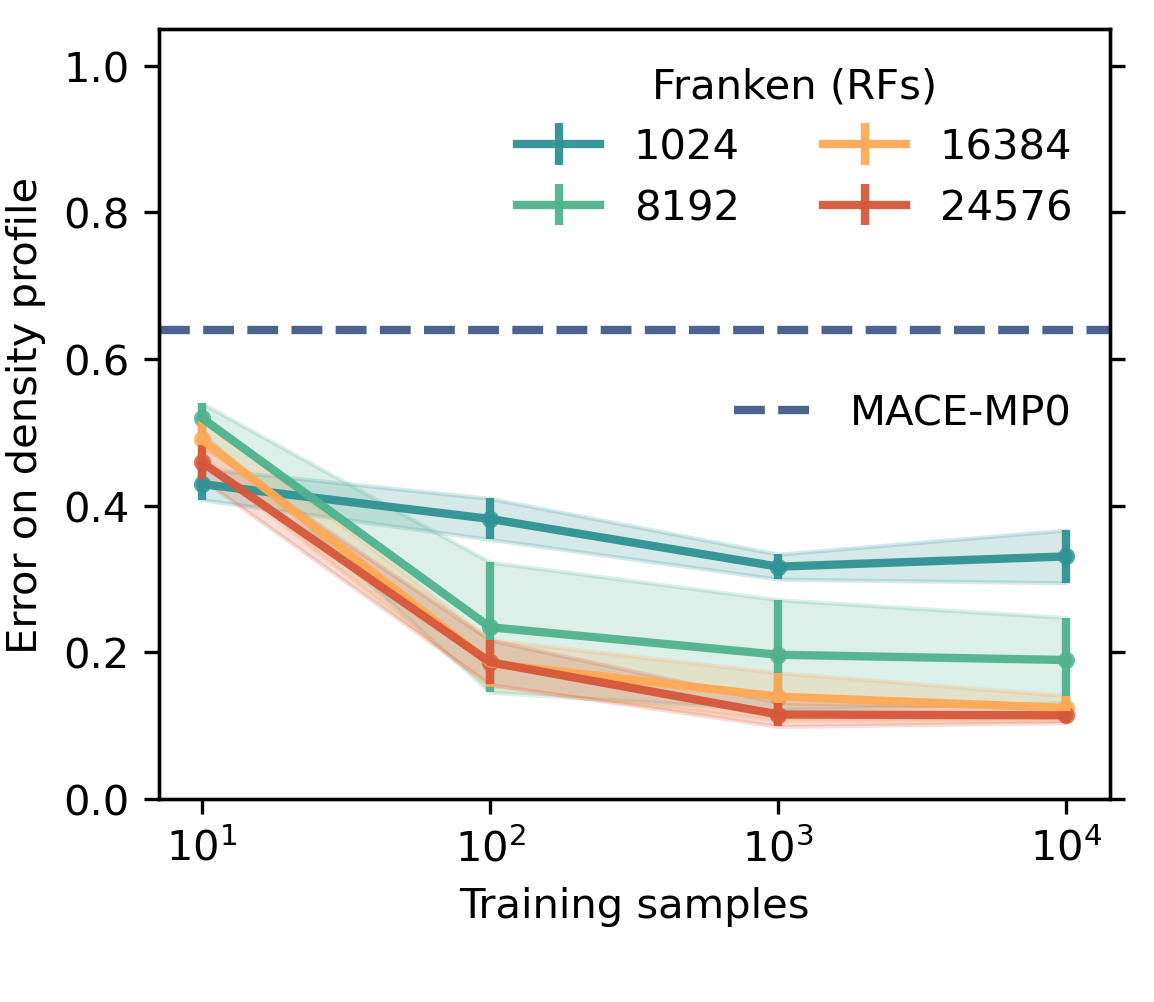}
    \caption{\textbf{Data-efficiency of MD observables}. Integrated error on the density profile with respect to the independent reference in \cref{fig:interface-2} for the zero-shot (MACE-MP0) simulation and the ones obtained with \algo~using different numbers of training samples. Shaded areas indicate standard deviations obtained from 5 independent replicas.  }
    \label{fig:SI-interface-sample-complexity}
\end{figure}
\end{document}